\documentclass[aps,prl,reprint,superscriptaddress,nofootinbib]{revtex4-2}
\usepackage{amsmath,amssymb,bm,mathtools,booktabs,graphicx,microtype}
\usepackage[colorlinks=true,allcolors=blue]{hyperref}
\newcommand{\ee}{\mathrm{e}}
\newcommand{\ii}{\mathrm{i}}

\newcommand{\var}{\operatorname{Var}}
\newcommand{\Tr}{\operatorname{Tr}}
\begin{document}

\makeatletter
\let\arxiv@addcontentsline\addcontentsline
\let\addcontentsline\@gobblethree
\makeatother

\title{Twist-controlled imaginary-time Mpemba effect in quantum spin systems}

\author{Zhi-Yao Ning}
\email{zyning39@gmail.com}
\affiliation{Institute of Solid-State Physics, University of Tokyo, Tokyo, 277-0882, Japan}

\author{Zi-Xiang Li}
\email{zixiangli@iphy.ac.cn}
\affiliation{Beijing National Laboratory for Condensed Matter Physics and Institute of Physics, Chinese Academy of Sciences, Beijing 100190, China}
\affiliation{University of Chinese Academy of Sciences, Beijing 100049, China}

\author{Naoki Kawashima}
\email{kawashima@issp.u-tokyo.ac.jp}
\affiliation{Institute of Solid-State Physics, University of Tokyo, Tokyo, 277-0882, Japan}
\affiliation{Trans-scale Quantum Science Institute, The University of Tokyo, Tokyo 113-0033, Japan}

\date{October 27, 2026}
\begin{abstract}
We construct a controllable imaginary-time Mpemba effect in quantum spin systems using spatially modulated local rotations without changing initial bipartite entanglement spectra and correlation magnitudes. This feature unambiguously isolates the redistribution of spectral weights as the sole driver of the anomalous relaxation. Using the compact-boson description, we reveal that these twists act as precise mode selectors: higher-harmonic twists exclusively populate high-energy standing waves, structurally bypassing slow-decaying infrared modes to accelerate convergence. We derive exact analytical crossing times for these reversals and demonstrate that intentionally reintroducing the slow-mode population systematically delays the ITME. This exact mechanism is validated beyond the fixed-point theory via calculations on interacting XXZ chains and ladders. Furthermore, a twelve-qubit simulation of variational quantum imaginary-time evolution reproduces the reversals and their controlled delay, providing a concrete route to realize this protocol in realistic quantum circuits. Ultimately, these results provide a strictly controllable and exact starting point for realizing and investigating the imaginary-time Mpemba effect, establishing a foundational paradigm to harness anomalous non-equilibrium relaxation in quantum ground-state preparations.
\end{abstract}
\maketitle

The classical Mpemba effect describes an anomalous relaxation phenomenon where an initially hotter system can freeze faster than a colder one~\cite{LuRaz2017,KlichEtAl2019}. Real-time quantum counterparts extend this reversal to the relaxation toward stationary states in dissipative systems~\cite{CarolloEtAl2021,KochsiekEtAl2022} and the restoration of broken symmetries following quenches or measurements~\cite{AresEtAl2023,RylandsEtAl2024,JoshiEtAl2024,DiGiulioEtAl2025}. These phenomena reveal how initial conditions redistribute spectral weights among the modes governing relaxation, thereby advancing our understanding of non-equilibrium phenomena in quantum systems. In this context, imaginary-time dynamics provides a clean spectral setting to study this reversal. It plays a fundamentally important role in modern computational physics, underlying quantum Monte Carlo methods~\cite{Sandvik2010,AssaadEvertz2008,LiYao2019Review,BeccaSorella2017,YuEtAl2026SignProblem}, tensor-network techniques~\cite{Schollwock2011,Orus2014,PaeckelEtAl2019}, and emerging quantum algorithms~\cite{McArdleEtAl2019,MottaEtAl2020}. Because normalized imaginary-time evolution suppresses excited components exponentially according to their energy gaps, an initially higher-energy state can overtake a lower-energy state as they converge toward a common ground state. This imaginary-time Mpemba effect (ITME)~\cite{ChangEtAl2026ITME,YuEtAl2026Pontus,GuoWang2026ITME,ZengEtAl2026ITME} directly reflects the initial excitation weights. Furthermore, it offers a promising framework to bridge ground-state properties with the rate of imaginary-time evolution~\cite{ChangEtAl2026ITME}, and, more crucially, provides a plausible route toward enhancing classical and quantum computations.

In quantum many-body systems, existing demonstrations of the ITME rely primarily on numerical simulations, often lacking precise control over the underlying mechanisms. Specifically, variables such as energy, correlations, and entanglement tend to change simultaneously across different initial states, obscuring the exact origin of the reversal. A theoretically designed ITME with precise controllability is therefore still lacking, yet it is essential for understanding how the fundamental features of imaginary-time evolution are encoded in the Hamiltonian and initial states. Furthermore, an exact design of a controlled ITME provides pivotal guidance for its realization in realistic quantum devices. Hence, it is valuable to propose an exact and controlled realization of the ITME, which not only holds crucial theoretical value but also paves the way for experimental implementation.

Twist operators, familiar from Lieb--Schultz--Mattis arguments and polarization diagnostics~\cite{LiebSchultzMattis1961,Tasaki2018,Resta1998,KobayashiEtAl2018,su2025identification}, offer a natural route to address this challenge. In this Letter, we apply zero-winding sinusoidal twists to a common ground state to selectively populate specific excitations, thereby proposing a general framework to realize the ITME in a controllable manner. In the compact-boson description, each twist displaces a selected standing-wave oscillator, yielding exact analytical crossing times for both energy and target infidelity. We introduce two controls to isolate the underlying mechanism: varying the amplitude at a fixed harmonic produces no reversal, whereas restoring the slow-mode weight systematically delays and eventually prevents it. We validate this theoretical construction using XXZ chains and extend it beyond the exactly solvable regime to the N\'eel phase, ferromagnets, and finite-width ladders. Furthermore, leveraging this designed approach, we propose a concrete scheme to realize the ITME in quantum circuits and perform noiseless simulations via a variational quantum imaginary-time evolution circuit to corroborate our findings.

\emph{Spatial twist perturbation.}--- We apply spatial twist $\hat{U}_m=\ee^{\ii\sum_j f_m(j)\hat{S}_j^z}$ to the ground state $|\psi_0\rangle$ of the quantum spin-$1/2$ XXZ chain in periodic boundary conditions
\begin{equation}
 H=J\sum_j(S_j^xS_{j+1}^x+S_j^yS_{j+1}^y+
 \Delta S_j^zS_{j+1}^z),\qquad J>0 ,
 \label{eq:xxz}
\end{equation}
at zero magnetization sector. Such a twist rotates the transverse spin components by a local phase factor $\langle S_i^+S_j^-\rangle_f=\ee^{\ii[f_m(j)-f_m(i)]}\langle S_i^+S_j^-\rangle_0$, leaving the longitudinal components unchanged. In this work we consider a sinusoidal pattern $f_m(j)=\alpha\sin(q_m j)$ with wavenumber $q_m=2\pi m/L$, amplitude $0<\alpha<1$ and harmonic index $m$.  Crucially, this twist serves as a highly controlled excitation protocol. Because $\hat{U}_m=\hat{U}_A\otimes \hat{U}_{\bar A}$ for any real-space bipartition $A\cup\bar A$, the reduced density matrix transforms as $\rho_A(f)=\hat{U}_A\rho_A(0)\hat{U}_A^\dagger$, preserving the Schmidt spectrum. Consequently, we isolate the initial spectral weight distribution as the sole driving mechanism for the ITME, decoupled from entanglement or correlation variations.

For $-1<\Delta<1$, the spin chain is a Luttinger liquid, and its low-energy physics is described by a compact boson
\begin{equation}
 H_{\rm LL}=\frac{v}{2\pi}\int_0^L dx\,
 \big[K(\partial_x\theta)^2+K^{-1}(\partial_x\phi)^2\big],
 \label{eq:ll}
\end{equation}
where $S^z(x)\simeq-\partial_x\phi/\pi$~\cite{Haldane1980,Haldane1981,Giamarchi2004}. Writing $\mu=\arccos\Delta$, Bethe ansatz fixes $K=\pi/[2(\pi-\mu)]$ and
$v=J\pi\sin\mu/(2\mu)$~\cite{Haldane1980,LutherPeschel1975}. The $\Delta=1$ limit has $K=1/2$, $v=\pi J/2$, with marginal corrections to the fixed-point description~\cite{Lukyanov1998}. For $\Delta>1$, the chain is gapped in the N\'eel phase, while at $\Delta=-1$ it is a ferromagnet with quadratic low-energy excitations.

\begin{figure}[!tb]
 \centering
 \includegraphics[width=\columnwidth]{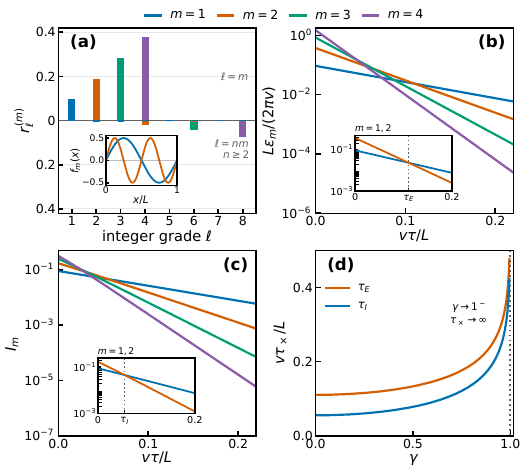}
 \caption{\label{fig:boson} Twisted states preparation and imaginary-time evolution in compact boson with $K=3/4$ and $\alpha=1/2$.
 (a) Target-normalized weights $r_\ell^{(m)}$ for different $m$ at integer grades $\ell$. Bars above zero show the corresponding state's weight at $\ell=m$; bars below show the other states' weights at each grade.
 Lower-left inset: twist profiles $f_m(x)$ for $m=1,2$.
 (b) Excess energy and (c) target infidelity; lower-left insets isolate $m_c=1,m_h=2$. Dotted lines in the insets mark the analytically determined crossing times $\tau_E/L=\ln2/(2\pi v)$ and $\tau_I/L=\ln2/(4\pi v)$.
 (d) $(m_h,m_c)=(2,1)$, restoring the slow harmonic to the hot state delays both crossings, which diverge logarithmically as $\gamma\to1^-$.}
\end{figure}

In the compact-boson language, the quadratic Hamiltonian~\eqref{eq:ll} describes independent harmonic oscillators for the nonzero Fourier modes, and the sinusoidal twist $\hat{U}_m=\ee^{\ii\alpha \hat{G}_m}$ selects a standing-wave combination of the degenerate $\pm q_m$ modes. The generator
$$\hat{G}_m=-\frac{1}{\pi}\int_0^L dx\,\sin (q_m x)\partial_x\hat{\phi}(x)=\sqrt{\frac{Km}{2}}(\hat{a}_m+\hat{a}_m^\dagger)$$
excites a coherent mode that satisfies $\hat{H}_m=\omega_m \hat{a}_m^\dagger \hat{a}_m$, with $\omega_m=vq_m$.
Therefore, the initial state $|\psi_m\rangle = \hat{U}_m|\psi_0\rangle$ is elegantly mapped to a coherent state $|\beta_m\rangle$ of the selected mode $m$, with displacement $\beta_m=\ii\alpha\sqrt{Km/2}$ and mean occupation $A_m=K\alpha^2m/2$. This mode-selective excitation provides the microscopic origin of the ITME. The target-normalized excitation weights follow a Poisson distribution $r_{nm}^{(m)} = A_m^n / n!$. Crucially, a twist with harmonic $m$ strictly populates energy levels associated with multiples of $m$ (i.e., $n \times m$), leaving all other modes unpopulated. As illustrated in Fig.~\ref{fig:boson}(a), choosing a higher harmonic (the ``hot'' state) deliberately bypasses the low-energy slow-decaying modes. Although it starts with a higher energy and larger infidelity, its components decay exponentially faster in imaginary time, ultimately overtaking the ``cold'' state.

\emph{Imaginary-time evolution.}---
We combine fixed-point-theory analyses of compact boson with matrix product states (MPS) calculations of XXZ spin chains. Using density-matrix renormalization group (DMRG) and time-dependent variational principle (TDVP)~\cite{White1992,HaegemanEtAl2016,FishmanEtAl2022}, we prepare and evolve the $m$-twisted states in imaginary time under $H_0=H-\langle H\rangle_0$ towards a common target $|\psi_0\rangle$:
\begin{equation}
 \begin{gathered}
 |\phi_m\rangle=U_m|\psi_0\rangle,\qquad
 |\psi_m(\tau)\rangle=
 \frac{\ee^{-\tau H_0}|\phi_m \rangle}
 {\|\ee^{-\tau H_0}|\phi_m \rangle\|}.
 \end{gathered}
 \label{eq:preparation}
\end{equation}
The initial target overlap is nonzero $\langle\psi_0|\psi_m(0)\rangle >0$ here. The excess energy $\varepsilon_m(\tau)=\langle \psi_m(\tau) | H_0 | \psi_m(\tau) \rangle$ and target infidelity $I_m=1-|\langle\psi_0|\psi_m(\tau)\rangle|^2$ show the distance from the target state.
A pair of twisted states, labelled as hot and cold, are prepared with different twists $(f_{m_h},f_{m_c})$ with $m_h>m_c$ and equal amplitude. The normalized differences of energy and infidelity
\begin{equation}
 \mathcal D_E(\tau)=
 \frac{\varepsilon_h(\tau)-\varepsilon_c(\tau)}
 {\varepsilon_h(0)-\varepsilon_c(0)},\qquad
 \mathcal D_I(\tau)=\frac{I_h(\tau)-I_c(\tau)}{I_h(0)-I_c(0)}
 \label{eq:differences}
\end{equation}
are initially positive, and the positive roots define the corresponding ITME.

\emph{ITME in harmonic oscillator.}--- We first study the compact boson analytically, where the twist excites a single harmonic mode.
Imaginary-time evolution reduces each harmonic mode individually by $\ee^{-\omega_m\tau}$:
\begin{equation}
 \begin{aligned}
 \varepsilon_m(\tau)&=\omega_m A_m\ee^{-2\omega_m\tau},\\
 I_m(\tau)&=1-\exp[-A_m\ee^{-2\omega_m\tau}].
 \end{aligned}
 \label{eq:boson}
\end{equation}
At equal amplitude, a higher harmonic starts with larger energy and infidelity but decays faster, producing the crossings in Figs.~\ref{fig:boson}(b,c). The crossing times balance the initial oscillator occupations against their decay rates. For a selected pair of harmonics $(m_h,m_c)$, substituting Eq.~\eqref{eq:boson} into Eq.~\eqref{eq:differences} and solving for the positive-time zeros of $\mathcal D_E(\tau)$ and $\mathcal D_I(\tau)$ yields
\begin{equation}
 \begin{aligned}
 \tau_E=\frac{\ln[\omega_{m_h}A_{m_h}/(\omega_{m_c}A_{m_c})]}{2(\omega_{m_h}-\omega_{m_c})},\,\, \tau_I=\frac{\ln(A_{m_h}/A_{m_c})} {2(\omega_{m_h}-\omega_{m_c})}.
 \end{aligned}
 \label{eq:roots}
\end{equation}
These expressions are exact at the Gaussian fixed point for the coherent oscillators in Eq.~\eqref{eq:boson}. At equal twist amplitude, $\omega_m\propto m/L$ and $A_m\propto m$, yielding
\begin{equation}
    \tau_E=2\tau_I=\frac{L\ln(m_h/m_c)}{2\pi v(m_h-m_c)},
    \label{eq:roots_equal}
\end{equation}
The crossing time is set by velocity $v$, while the population follows $r_1c/\alpha^2=K(\Delta)/2$.

Two controls isolate the mechanism. First, at fixed harmonic $m$, increasing the amplitude to $\alpha_h>\alpha_c>0$ raises the initial energy and infidelity. Both observables retain their ordering and no ITME occurs since the decay rates are the same.
Second, retain the cold profile $f_c=\alpha\sin(2\pi m_c x/L)$ and add its slow harmonic to the hot profile: $f_h=\alpha[\sin(2\pi m_h x/L)+\gamma\sin(2\pi m_c x/L)]$. The restored slow-mode occupation is $\gamma^2A_{m_c}$, delaying both crossings despite increasing the hot state's initial energy and infidelity. For $m_h>m_c>0$ and $0\leq\gamma<1$, solving Eq.~\eqref{eq:differences} gives the crossing times
\begin{equation}
 \begin{aligned}
 \tau_E(\gamma)&=\frac{\ln[(m_h/m_c)^2/(1-\gamma^2)]}
 {2(\omega_{m_h}-\omega_{m_c})},\\
 \tau_I(\gamma)&=\frac{\ln[(m_h/m_c)/(1-\gamma^2)]}
 {2(\omega_{m_h}-\omega_{m_c})}.
 \end{aligned}
 \label{eq:contamination}
\end{equation}
Both crossing times increase and diverge as $\gamma\to1^-$ [Fig.~\ref{fig:boson}(d)]. At $\gamma\geq1$, the hot state's slow-mode weight equals or exceeds the cold state's, preventing either reversal. Thus ITME depends on modal composition and suppression of slow-mode weight, not on higher initial energy alone.

\begin{figure}[!tb]
 \centering
 \includegraphics[width=\columnwidth]{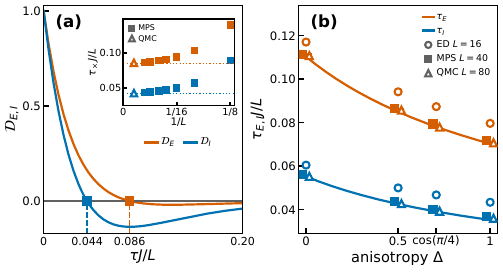}
 \caption{\label{fig:critical}ITME in the critical XXZ chain.
 (a) Normalized energy (orange) and infidelity (blue) differences at $\Delta=1/2$, from periodic MPS at $L=40$.
 Inset: crossing times versus inverse size (squares); dotted lines show the compact-boson limits.
 (b) Crossing times versus anisotropy. Solid curves are the compact-boson formula of Eq.~\eqref{eq:roots} with Bethe-ansatz velocity; open circles, open triangles, and filled squares denote ED, QMC, and MPS, respectively, for four representative anisotropies. MPS and QMC markers are slightly displaced horizontally for visibility.  }
\end{figure}

\emph{ITME in XXZ Luttinger liquid.}--- We calculate the energy and infidelity differences in Eq.~\eqref{eq:differences} for the XXZ chain by DMRG and TDVP. In $|\Delta|<1$, the chain is in the Luttinger liquid phase, and can be compared with the compact-boson analysis.
Fig.~\ref{fig:critical}(a) shows the MPS results of $(m_c,m_h)=(1,2)$ pair at $\Delta=1/2$, where both differences cross zero at positive times, confirming the ITME. The inset shows the crossing times versus inverse system size, approaching the compact-boson predictions from Eq.~\eqref{eq:roots} as $L$ increases.
Both crossing times decrease with anisotropy, following the increasing mode velocity as compact-boson analysis [Fig.~\ref{fig:critical}(b)].
Directed-loop quantum Monte Carlo (QMC) provides independent large-scale verification of both reversals at $L=80$~\cite{SyljuasenSandvik2002,DorneichTroyer2001} using $\beta J=4L$ (error estimation see Supplemental Material).

Microscopically, this twist-induced ITME is produced by two exact lattice properties: geometric energy injection and momentum selection.  The twist injects energy by rotating the transverse bond  $C_\perp=\langle S_j^xS_{j+1}^x+S_j^yS_{j+1}^y\rangle_0$ with angle on each bond $\delta f_j=f_m(j+1)-f_m(j)$:
\begin{equation}
 \varepsilon_m(0)=JC_\perp\sum_j(\cos\delta f_j-1)\simeq-\frac{JC_\perp\pi^2\alpha^2m^2}{L}.
 \label{eq:injection}
\end{equation}
The second formula holds for $q_m\ll1$ in the continuous-field limit. Since $C_\perp<0$, higher harmonics inject more energy through steeper phase gradients. Crucially, this energy injection mechanism relies purely on short-range correlations rather than low-energy field theory, making it applicable even beyond the compact-boson regime ($\Delta \ge 1$). The twist also obeys a lattice momentum-selection rule. For $d=\gcd(m,L)$, $f_m(j+L/d)=f_m(j)$ implies $[\hat U_m,\hat T^{L/d}]=0$, where $\hat T$ translates by one site. If the target momentum is $k_0=2\pi\nu_0/L$, then $\hat{U}_m$ only populates momentum indices $\nu$ satisfying $k_\nu-k_0=2\pi(\nu-\nu_0)/L\in(2\pi d/L)\mathbb Z$. In particular, $\hat{U}_{m=2}$ excludes odd relative momentum indices on an even ring, as shown in Fig.~\ref{fig:boson}(a). By structurally eliminating the slowest infrared relaxations, this selection rule ensures the accelerated convergence of the higher harmonic state.

\begin{figure}[!tb]
 \centering
 \includegraphics[width=0.95\columnwidth]{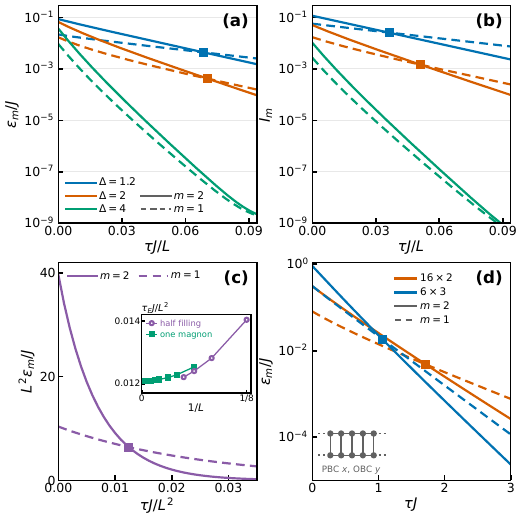}
 \caption{\label{fig:beyond}Beyond the linear compact boson, calculated by MPS at $\alpha=1/2$. Solid and dashed curves denote hot $m_h=2$ and cold $m_c=1$ trajectories; filled squares mark the crossings respectively.
 (a) Excess energy and (b) target infidelity in the N\'eel phase at $L=32$. Filled squares mark resolved crossings at $\Delta=1.2,2$, while no resolved crossing for $\Delta=4$ up to numerical resolution.
 (c) Energy crossing at $\Delta=-1$ in the half-filled $n=L/2$, $L=16$ sector. Inset: $\tau_EJ/L^2$ for half-filled results and the distinct one-magnon ($n=1$) sequence.
 (d) Excess-energy crossings on $16\times2$ (orange) and $6\times3$ (blue) ladders at $\Delta=1/2$, periodic along $x$ and open along $y$. }
\end{figure}

\emph{Beyond compact bosons.}---For $\Delta>1$, the harmonic oscillator is no longer the low-energy description, yet the same preparation can still produce ITME [Fig.~\ref{fig:beyond}(a,b)]. Deep in the N\'eel phase, the transverse fluctuations of the spins decrease, suppressing the energy injected by the twist.
In the Ising and long-wavelength limits the leading injected energy scales as $\varepsilon_m(0)\sim J\pi^2\alpha^2m^2/(2\Delta L)$, while the domain-wall excitations carrying the weight cost order $J\Delta$. The twisted states therefore start closer to the target and relax faster, and the cold-state deviation can fall below the numerical resolution before a crossing becomes resolvable. At $L=32$, crossings are still resolvable for $\Delta=1.2,2$. But for $\Delta=4$, the excess energy and infidelity of cold state reach numerical precision $10^{-8}$ before overtaken by the hot state, so no resolved crossing is found within the numerical resolution.

The $\Delta=-1$ point provides the $z=2$ comparison. A staggered rotation $W=\prod_{j\,{\rm odd}}(2S_j^z)$ maps Eq.~\eqref{eq:xxz} to $H_F=-J\sum_j\bm S_j\cdot\bm S_{j+1}$ and commutes with $U_f$~\cite{AlbaHaqueLauchli2012}. Its sector-specific target is $W^\dagger|D_n\rangle$, where $|D_n\rangle\propto(S^-_{\rm tot})^n|F\rangle$ is a symmetric Dicke state and $|F\rangle$ is fully spin-up~\cite{Dicke1954}. Here $n$ counts down spins and $n/L$ is the filling. Figure~\ref{fig:beyond}(c) shows the many-body half-filled sector $n=L/2$ in its main panel, while the inset also compares the one-magnon sector $n=1$. A weak sinusoidal twist populates a spin-wave mode with gap $e_m=J[1-\cos(2\pi m/L)]\simeq2\pi^2Jm^2/L^2$. Its leading excited population and injected energy are
\begin{equation}
 w_m\simeq\frac{\alpha^2n(L-n)}{2(L-1)},\quad \varepsilon_m(0)\simeq e_mw_m.
\label{eq:injection_ferro}
\end{equation}
Equal amplitudes give $w_h=w_c$ at this order, unlike $A_m\propto m$ in the compact boson. The analogous leading-mode balance gives $\tau_E=\ln(e_h/e_c)/[2(e_h-e_c)]\propto L^2$, but no positive-time infidelity reversal. The half-filled trajectories and one-magnon sequence in Fig.~\ref{fig:beyond}(c) are consistent with this quadratic energy-crossing scale. The fully polarized sectors $n=0,L$ are null cases since the twist produces only a trivial phase.

More generally, when a weak twist excites population $w_m$ concentrated at a gap $e_m$, we can write down the injected energy $\varepsilon_m\simeq e_mw_m\ee^{-2e_m\tau}$ and $I_m\simeq w_m\ee^{-2e_m\tau}$. Their crossing times follow the balance in Eq.~\eqref{eq:roots}, with $\omega_m,A_m$ replaced by $e_m,w_m$, respectively. If $e(q)\propto|q|^z$, then $e_m\propto m^zL^{-z}$ at fixed harmonic index. Provided $w_h/w_c$ is size independent and a positive crossing exists, $\tau_\times\propto L^z$: linearly for the $z=1$ Luttinger liquid, quadratically for the $z=2$ ferromagnet, and as a higher power for larger $z$. This connects to the potential efficiency gains associated with larger $z$~\cite{ChangEtAl2026ITME}, but separates the inverse-gap timescale from the spectral-weight condition for ITME.

The twist-perturbed ITME need not be restricted to a chain. With $f(x,y)=\alpha\sin(2\pi mx/L_x)$ on a finite-width XXZ ladder, both energy and infidelity reverse for the two- and three-leg geometries in Fig.~\ref{fig:beyond}(d). These numerical results demonstrate that the twist-controlled ITME can be applied beyond the compact-boson regime and the one-dimensional chain, and can be realized in a wide range of quantum spin systems.

\begin{figure}[!tb]
 \centering
 \includegraphics[width=\columnwidth]{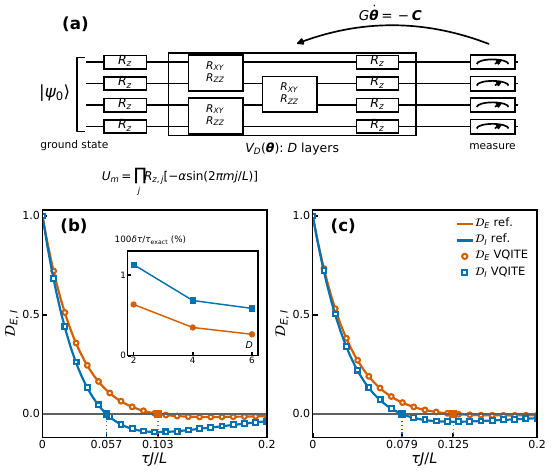}
 \caption{\label{fig:circuit}A circuit realization by variational quantum  imaginary-time evolution (VQITE).
 (a) A diagrammatic illustration of the VQITE circuit.
 (b) At $L=12$, $\Delta=\alpha=1/2$, normalized differences from the MPS reference (lines) and $D=6$, $M_a=2$ VQITE (open markers). Inset: signed percentage errors of the crossing times at depths $D$, where $\delta\tau=\tau_{\rm VQITE}-\tau_{\rm exact}$.
 (c) Slow-mode admixture $\gamma=0.5$ at the same parameters, comparing the exact reference (lines) and VQITE (open markers). The legend in (c) applies to both lower panels. Orange and blue denote energy and infidelity, respectively; squares and dotted guides mark the reference crossing times.}
\end{figure}

\emph{Variational quantum imaginary-time evolution.}---Quantum imaginary-time methods provide a framework to translate nonunitary relaxation into measurable circuit operations~\cite{MottaEtAl2020,McArdleEtAl2019,ZhangYin2024}.
Critically, our theoretically designed twist perturbation translates natively to quantum hardware as a simple product of programmable local phase gates, $U_m=\prod_jR_{z,j}[-\alpha\sin(2\pi mj/L)]$.  The ground state $|\psi_0\rangle=W_0|0\cdots0\rangle$ constructed by the ground-state preparation circuit $W_0$ from the initialized qubits $|0\cdots0\rangle$. The subsequent state is approximated by $V_D(\bm\theta)U_m|\psi_0\rangle$, where $V_D$ is a magnetization-conserving brick-wall circuit with $D$ layers and a truncated spatial Fourier parametrization [Fig.~\ref{fig:circuit}(a)]. Its two-qubit blocks combine exchange $R_{XY}$ and $R_{ZZ}$ gates, interleaved with local $R_z$ rotations. The meter symbols denote measurement settings for the metric $G$, energy gradient $\bm C$, and relaxation observables. Classical feedback updates the circuit angles according to the McLachlan equation $G\dot{\bm\theta}=-\bm C$, which projects the imaginary-time direction onto this variational tangent space~\cite{McArdleEtAl2019,YuanEtAl2019}.

We simulate the twelve-qubit VQITE circuit using TensorCircuit with its JAX backend~\cite{ZhangEtAlTensorCircuit2023}, reproducing ITME for both energy and infidelity and comparing with the MPS references [Fig.~\ref{fig:circuit}(b)]. At bandwidth $M_a=2$, increasing depth from $D=2$ to $6$ reduces the energy-root error from $0.64\%$ to $0.27\%$ and the infidelity-root error from $1.13\%$ to $0.59\%$. Furthermore, the variational circuit successfully captures our finely tuned physical control: injecting a slow-mode admixture ($\gamma=0.5$) dynamically delays both crossings exactly as predicted [Fig.~\ref{fig:circuit}(c)].

\emph{Discussion.}---
Spatially modulated twists provide a theoretically exact and broadly applicable framework to study the ITME with rigorously controlled initializations. By strictly preserving the initial bipartite entanglement spectrum and correlation envelopes, our protocol isolates spectral weight redistribution as the definitive driver of the anomalous relaxation. The compact boson theory offers a quantitative backbone for this mechanism, revealing how harmonic selection dictates decay rates and crossing times. Our XXZ-chain and finite-width ladder calculations further demonstrate that this mode-selective principle is not limited to linearly dispersing oscillators, but extends robustly to interacting lattice models. Since the preparation and evolution conserve total magnetization, these energy and target-infidelity reversals are distinct from restoration of a broken U(1) symmetry measured by entanglement asymmetry~\cite{AresEtAl2023}.

The local-$R_z$ preparation and noiseless VQITE results suggest a concrete route to testing twist-controlled ITME on quantum devices~\cite{MottaEtAl2020,McArdleEtAl2019,ZhangYin2024}. Experiments can apply selected phase textures to a common target and track energy and target fidelity, with same-harmonic amplitude and slow-mode-admixture controls isolating the spectral mechanism. Such hardware demonstrations would establish the robustness of mode-selective relaxations against preparation and gate errors, evaluating their potential as a deterministic strategy to bypass spectral slowing-down and reduce the resource overhead for reaching fixed target accuracies.

The same mode-selection principle suggests analogies in other quantum magnets. Replacing $S_j^z$ by the density $n_j$ gives a charge-phase twist in fermionic Luttinger liquids~\cite{Haldane1981,Giamarchi2004}, while Cartan-generator twists extend the question to SU($N$) chains~\cite{Sutherland1975,Affleck1988,su2024exact}.

Normalized imaginary time also admits a conditional non-Hermitian interpretation, connecting our construction to monitored quantum dynamics~\cite{DalibardCastinMolmer1992}. For $H_{\mathrm{eff}}=-\ii H_0$, the normalized no-jump trajectory coincides with the imaginary-time state, and its conditional jump rate satisfies $\lambda_f=2\varepsilon_f$. An energy reversal is thus a crossing of conditional jump rates. The coherent states generated in the compact-boson description offer a simple dissipative analogue: under independent zero-temperature linear loss, quantum jumps leave the normalized coherent state unchanged~\cite{Longhi2024BosonicMpemba}. This motivates exploring how spatial phase control can shape relaxation in open quantum magnets beyond the coherent-state limit. For the present twist construction, the symmetry and reservoir conditions for suppressing slow dissipative modes, and the effects of quantum jumps on the resulting relaxation reversals remain open questions~\cite{MingantiEtAl2019,WeiEtAl2026SymmetryMpemba}.

\begin{acknowledgments}
We thank Shuai Yin, Hang Su, Yuan Yao, Hui Shao, Jun Takahashi and Myles Scollon for valuable discussions and comments. Z.-X.L. is supported by the National Natural Science Foundation of China under Grant No. 12347107 and No. 12474146, Beijing Natural Science Foundation under Grant No. JR25007 and the New Cornerstone Investigator Program. N.K. is supported by MEXT with JSPS KAKENHI Grant No.23K25789.
\end{acknowledgments}

\makeatletter
\let\addcontentsline\@gobblethree
\bibliography{references}
\let\addcontentsline\arxiv@addcontentsline
\makeatother
\clearpage
\onecolumngrid
\setcounter{equation}{0}
\setcounter{figure}{0}
\setcounter{table}{0}
\setcounter{section}{0}
\setcounter{subsection}{0}
\renewcommand{\theequation}{S\arabic{equation}}
\renewcommand{\thefigure}{S\arabic{figure}}
\renewcommand{\thetable}{S\arabic{table}}
\renewcommand{\theHequation}{supp.\arabic{equation}}
\renewcommand{\theHfigure}{supp.\arabic{figure}}
\renewcommand{\theHtable}{supp.\arabic{table}}
\renewcommand{\theHsection}{supp.\arabic{section}}
\renewcommand{\theHsubsection}{supp.\arabic{section}.\arabic{subsection}}
\setcounter{secnumdepth}{2}
\setcounter{tocdepth}{2}
\pdfbookmark[0]{Supplemental Material}{supplement-start}
\begin{center}
{\large\bfseries Supplemental Material}\\[0.5em]
{\bfseries Twist-controlled imaginary-time Mpemba effect in quantum spin systems}
\end{center}
\medskip

\noindent
This Supplemental Material collects the derivations and numerical details that support the Letter. Section~\ref{sec:spectral} summarizes the spectral description of normalized imaginary-time evolution and the properties of the twist operator. Section~\ref{sec:boson} derives the exact solution at the Gaussian (compact-boson) fixed point, including the two mechanism controls. Section~\ref{sec:critical} turns to the microscopic XXZ chain: the critical regime, the gapped N\'eel regime, the ferromagnetic endpoint, and finite-width ladders. Sections~\ref{sec:qmc} and \ref{sec:vqite} describe the quantum Monte Carlo (QMC) and variational quantum imaginary-time evolution (VQITE) calculations, and Sec.~\ref{sec:nonhermitian} discusses the relation to monitored quantum dynamics.

Throughout, $\hbar=a=1$, $J>0$ sets the energy unit, and $\varepsilon=\langle H\rangle-E_0$ denotes the \emph{total} excess energy rather than the energy per site. Unless stated otherwise, we consider a periodic chain with the target state in the $S^z_{\rm tot}=0$ sector, twist amplitude $\alpha=1/2$, and the harmonic pair $(m_h,m_c)=(2,1)$. Imaginary time is measured in units of $L/J$ for the critical chain, $L^2/J$ at the ferromagnetic endpoint, and $1/J$ for gapped chains and ladders.
{\footnotesize\linespread{0.96}\selectfont\tableofcontents}
\clearpage

\section{Normalized imaginary-time evolution and twisted states}
\label{sec:spectral}
\subsection{Generating function and observables}
Let $|\psi_0\rangle$ be the unique ground state in the symmetry sector of interest, with energy $E_0$, and let $P_\nu$ project onto the eigenspace with excitation energy $e_\nu=E_\nu-E_0$ (grouping degenerate levels together). A prepared state $|\phi_f\rangle=U_f|\psi_0\rangle$ has spectral weights $p_{\nu,f}=\langle\phi_f|P_\nu|\phi_f\rangle$, and we assume a nonzero target overlap, $p_{0,f}>0$. Its normalized imaginary-time evolution is generated by the norm function $Z_f$,
\begin{equation}
 Z_f(s)=\sum_\nu p_{\nu,f}\ee^{-s e_\nu},\qquad
 |\psi_f(\tau)\rangle=
 \frac{\ee^{-\tau(H-E_0)}|\phi_f\rangle}{\sqrt{Z_f(2\tau)}}.
 \label{eq:S-Z}
\end{equation}
Since the norm contains two propagators, all observables are derivatives of $\ln Z_f$ evaluated at $s=2\tau$:
\begin{equation}
 \varepsilon_f(\tau)=-\partial_s\ln Z_f(s)\big|_{s=2\tau},\qquad
 \sigma_f^2(\tau)=\partial_s^2\ln Z_f(s)\big|_{s=2\tau},\qquad
 F_f(\tau)=1-I_f(\tau)=\frac{p_{0,f}}{Z_f(2\tau)}.
 \label{eq:S-observables}
\end{equation}
Here $\sigma_f^2$ is the energy variance and $F_f$ the fidelity to the common target. It is convenient to measure all weights relative to the target, $r_{\nu,f}=p_{\nu,f}/p_{0,f}$, and to define $R_f(\tau)=\sum_{\nu>0}r_{\nu,f}\ee^{-2e_\nu\tau}$. Then
\begin{equation}
 I_f=\frac{R_f}{1+R_f},\qquad
 \varepsilon_f=
 \frac{\sum_{\nu>0}e_\nu r_{\nu,f}\ee^{-2e_\nu\tau}}{1+R_f},\qquad
 D_f\equiv-\ln F_f=\ln(1+R_f).
 \label{eq:S-ratios}
\end{equation}
Both $I_f$ and the fidelity distance $D_f$ are monotonic functions of $R_f$, so they cross at the same times; we quote infidelity crossings, which apply equally to $D_f$.

\subsection{Cooling and the condition for a reversal}
From $\partial_\tau|\psi\rangle=-(H-\langle H\rangle)|\psi\rangle$ one obtains
\begin{equation}
 \partial_\tau\varepsilon=-2\sigma^2,\qquad
 \partial_\tau I=-2\varepsilon(1-I).
 \label{eq:S-cooling}
\end{equation}
Each state therefore cools monotonically, at a rate set by its own energy variance. As in the Letter, we compare two states through the normalized differences $\mathcal D_E=\delta\varepsilon(\tau)/\delta\varepsilon(0)$ and $\mathcal D_I=\delta I(\tau)/\delta I(0)$, with $\delta\varepsilon=\varepsilon_h-\varepsilon_c$ and $\delta I=I_h-I_c$; their positive zeros define the crossing times $\tau_E$ and $\tau_I$.

At late times both differences are dominated by the lowest excitation energy $e_*$ at which the target-normalized weights of the two states differ, and their sign is that of $r_{*,h}-r_{*,c}$~\cite{ZengEtAl2026ITME}. A reversal therefore persists if the hot state has the smaller weight in this slowest differing channel. Conversely, a first crossing can be followed by a recrossing when a still slower channel, more strongly populated in the hot state, takes over at late times.

\subsection{The twist operator: entanglement spectra and correlation envelopes}
\label{sec:twist}
The twist $U_f=\prod_j\ee^{\ii f_jS_j^z}$ is a product of on-site rotations. Although its form recalls the Lieb--Schultz--Mattis twist~\cite{LiebSchultzMattis1961,Tasaki2018}, the periodic, zero-winding profiles used here act as a preparation pulse rather than a persistent flux: the subsequent evolution uses the original Hamiltonian with unchanged boundary conditions. For any real-space bipartition $A\cup\bar A$ it factorizes as $U_f=U_A\otimes U_{\bar A}$, so that
\begin{equation}
 \rho_A(f)=\Tr_{\bar A}
 [U_f|\psi_0\rangle\langle\psi_0|U_f^\dagger]
 =U_A\rho_A(0)U_A^\dagger.
 \label{eq:S-schmidt}
\end{equation}
The reduced density matrix changes only by a unitary conjugation, and all entanglement entropies are unchanged, also for disconnected regions $A$.

The twist rotates the transverse spin components, $U_f^\dagger S_j^\pm U_f=\ee^{\mp\ii f_j}S_j^\pm$, so that
\begin{equation}
 \langle S_i^zS_j^z\rangle_f=\langle S_i^zS_j^z\rangle_0,
 \qquad
 \langle S_i^+S_j^-\rangle_f=
 \ee^{\ii(f_j-f_i)}\langle S_i^+S_j^-\rangle_0.
 \label{eq:S-correlations}
\end{equation}
Longitudinal correlations are unchanged. For a $U(1)$-symmetric target, whose transverse one-point functions vanish, the magnitude of the transverse correlations is also unchanged, while their phase acquires the spatial texture of $f$. The twisted states thus share entanglement spectra and correlation envelopes with the target at $\tau=0$, although these quantities evolve differently under the nonunitary dynamics. The energy cost of the twist and the lattice momentum selection it imposes are derived in Sec.~\ref{sec:critical-injection}.

\section{Compact boson: harmonic selection and exact crossing times}
\label{sec:boson}
\subsection{The selected standing-wave oscillator}
We use the compact-boson Hamiltonian
\begin{equation}
 H_{\rm LL}=\frac{v}{2\pi}\int_0^Ldx
 [K(\partial_x\theta)^2+K^{-1}(\partial_x\phi)^2],
 \qquad \rho(x)=-\frac1\pi\partial_x\phi(x),
 \label{eq:S-LL}
\end{equation}
where $\rho$ is the smooth part of the longitudinal spin density; the charge and winding zero modes are fixed. The Fourier components $\rho_q=\int_0^Ldx\,\ee^{-\ii qx}\rho(x)$ of the nonzero modes are independent oscillators~\cite{Haldane1980,Haldane1981,Giamarchi2004},
\begin{equation}
 \rho_q=\sqrt{\frac{KL|q|}{2\pi}}(b_q+b_{-q}^\dagger),\qquad
 H_{\pm q}=v|q|(b_q^\dagger b_q+b_{-q}^\dagger b_{-q}),
 \label{eq:S-densityosc}
\end{equation}
with the vacuum energy absorbed into $E_0$. The sinusoidal twist couples to $G_m=\int dx\,\sin(q_mx)\rho(x)=(\rho_{-q_m}-\rho_{q_m})/(2\ii)$ with $q_m=2\pi m/L$. In terms of the standing-wave combinations $a_m=\ii(b_{q_m}-b_{-q_m})/\sqrt2$ and $c_m=(b_{q_m}+b_{-q_m})/\sqrt2$,
\begin{equation}
 G_m=\sqrt{\frac{Km}{2}}(a_m+a_m^\dagger),\qquad
 H_{\pm q_m}=\omega_m(a_m^\dagger a_m+c_m^\dagger c_m),
 \qquad \omega_m=vq_m.
 \label{eq:S-standing}
\end{equation}
The twist therefore addresses a single standing-wave oscillator rather than one chiral mode. Consistently, the static structure factor $S^{zz}(q)=K|q|/(2\pi)$ gives $\langle G_m^2\rangle=LS^{zz}(q_m)/2=Km/2$.

\subsection{Coherent displacement and Poissonian grades}
With $X_m=(a_m+a_m^\dagger)/\sqrt2$, the twist $U_m=\ee^{\ii\alpha G_m}=\ee^{\ii\alpha\sqrt{Km}X_m}$ is a momentum displacement of oscillator $a_m$. It produces a coherent state,
\begin{equation}
 U_m|\psi_0\rangle=|\beta_m\rangle,
 \qquad \beta_m=\ii\alpha\sqrt{Km/2},\qquad
 A_m=|\beta_m|^2=K\alpha^2m/2,
 \label{eq:S-coherent}
\end{equation}
while all other oscillators remain in their vacua and the winding is unchanged. The mean occupation $A_m$ grows linearly with the harmonic index. Since the oscillator quanta have energy $\omega_m=m\omega_1$, the $n$-quantum component lies at the energy grade $E-E_0=\ell\omega_1$ with $\ell=nm$. The Poisson distribution of $n$ then gives
\begin{equation}
 p_{nm}^{(m)}=\ee^{-A_m}\frac{A_m^n}{n!},\qquad
 r_\ell^{(m)}=\frac{p_\ell^{(m)}}{p_0^{(m)}}=
 \begin{cases}A_m^{\ell/m}/(\ell/m)!,&m\mid\ell,\\
 0,&m\nmid\ell,\end{cases}
 \label{eq:S-gradeweights}
\end{equation}
with $r_0=1$ and $\sum_\ell r_\ell=\ee^{A_m}$. A twist with harmonic $m$ thus leaves all grades below $\ell=m$ empty. These are the weights shown in Letter Fig.~1(a), where the downward bars are drawn for clarity but represent positive weights. This Gaussian selection is stronger than the exact lattice momentum rule derived in Sec.~\ref{sec:critical-injection}.

\subsection{Imaginary-time filtering and crossing times}
Summing the Poisson series gives
\begin{equation}
 Z_m(s)=\exp\{A_m[\ee^{-s\omega_m}-1]\},\qquad
 \frac{\ee^{-\tau H_m}|\beta_m\rangle}
 {\|\ee^{-\tau H_m}|\beta_m\rangle\|}
 =|\beta_m\ee^{-\omega_m\tau}\rangle,
 \label{eq:S-Zcoherent}
\end{equation}
i.e., normalized imaginary-time evolution keeps the state coherent and shrinks its amplitude at the oscillator frequency. Hence
\begin{equation}
 D_m(\tau)=A_m\ee^{-2\omega_m\tau},\qquad
 \varepsilon_m(\tau)=\omega_mD_m(\tau),\qquad
 I_m(\tau)=1-\ee^{-D_m(\tau)}.
 \label{eq:S-boson}
\end{equation}
The energy and $D_m$ decay exactly at the rate $2\omega_m$. For $\omega_h>\omega_c$, equating the energies and, separately, the fidelity distances of the two states yields
\begin{equation}
 \tau_E=\frac{\ln(\omega_hA_h/\omega_cA_c)}
 {2(\omega_h-\omega_c)},\qquad
 \tau_I=\frac{\ln(A_h/A_c)}{2(\omega_h-\omega_c)}.
 \label{eq:S-generalroots}
\end{equation}
A crossing occurs when the corresponding numerator is positive, i.e., when the faster-decaying state starts with the larger energy or infidelity. At equal amplitude, $A_h/A_c=\omega_h/\omega_c=m_h/m_c$, and
\begin{equation}
 \tau_E=2\tau_I=
 \frac{L\ln(m_h/m_c)}{2\pi v(m_h-m_c)}.
 \label{eq:S-equalroots}
\end{equation}
The common amplitude $\alpha$ sets the size of the signal but cancels from the crossing times, which depend only on the velocity and scale linearly with $L$. The factor of two between $\tau_E$ and $\tau_I$ follows from the linear dispersion together with $A_m\propto m$ and is not universal.

\subsection{Two controls: amplitude at fixed harmonic and slow-mode admixture}
At fixed harmonic, a larger amplitude $\alpha_h>\alpha_c$ only rescales the occupation, so that $\varepsilon_h/\varepsilon_c=D_h/D_c=\alpha_h^2/\alpha_c^2>1$ at all times. Both states decay at the same rate, so neither ordering reverses although the hot state starts with higher excited energy.

To test the role of the slow mode directly, we add the cold harmonic to the hot profile, $f_c=\alpha\sin(q_{m_c}x)$ and $f_h=\alpha[\sin(q_{m_h}x)+\gamma\sin(q_{m_c}x)]$. Since the oscillators are independent, the hot state acquires the slow occupation $\gamma^2A_c$, and
\begin{align}
 D_h-D_c&=A_h\ee^{-2\omega_h\tau}
 -(1-\gamma^2)A_c\ee^{-2\omega_c\tau},\\
 \varepsilon_h-\varepsilon_c&=
 \omega_h A_h\ee^{-2\omega_h\tau}
 -(1-\gamma^2)\omega_c A_c\ee^{-2\omega_c\tau}.
 \label{eq:S-contaminationdiff}
\end{align}
At equal amplitude, for any $m_h>m_c>0$,
\begin{equation}
 \tau_E(\gamma)=
 \frac{\ln[(m_h/m_c)^2/(1-\gamma^2)]}{2(\omega_h-\omega_c)},
 \qquad
 \tau_I(\gamma)=
 \frac{\ln[(m_h/m_c)/(1-\gamma^2)]}{2(\omega_h-\omega_c)},
 \quad 0\leq\gamma<1.
 \label{eq:S-contamination}
\end{equation}
Restoring the slow mode raises the initial energy of the hot state, yet delays both crossings; they diverge logarithmically as $\gamma\to1^-$ while their separation stays fixed at $\ln(m_h/m_c)/[2(\omega_h-\omega_c)]$. For $\gamma\geq1$ the hot state has at least as much slow-mode weight as the cold one, and neither ordering reverses.

\begin{figure}[tb]
 \centering
 \includegraphics[width=.9\textwidth]{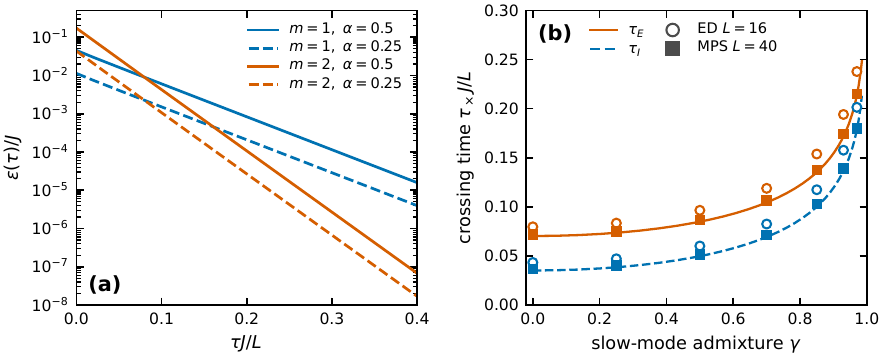}
\caption{\label{fig:S-controls}Mechanism controls in the Heisenberg chain ($\Delta=1$). (a) Excess energy for the same harmonic at two amplitudes ($L=16$, ED). On the logarithmic scale the curves for equal $m$ are parallel, i.e., they decay at the same rate and never cross; curves of different $m$ do cross. (b) Crossing times for $f_h=\alpha[\sin(q_2x)+\gamma\sin(q_1x)]$ and $f_c=\alpha\sin(q_1x)$ with $\alpha=1/2$, from ED at $L=16$ (open circles) and MPS at $L=40$ (filled squares), compared with the Gaussian result of Eq.~\eqref{eq:S-contamination} (lines). Numerical uncertainties are smaller than the symbols.}
\end{figure}

Figure~\ref{fig:S-controls} tests both controls in the Heisenberg chain, where the compact-boson description holds only at low energy. The same-harmonic curves remain parallel on a logarithmic scale throughout the window, so a larger amplitude alone never produces a reversal [Fig.~\ref{fig:S-controls}(a)]. With slow-mode admixture, both crossing times grow with $\gamma$ [Fig.~\ref{fig:S-controls}(b)], as predicted by Eq.~\eqref{eq:S-contamination}.

\section{XXZ chains: microscopic realization}
\label{sec:critical}
\subsection{Twist-injected energy and momentum selection}
\label{sec:critical-injection}
The energy injected by the twist follows from rotating the XXZ bonds. With $B_j=S_j^xS_{j+1}^x+S_j^yS_{j+1}^y$, $\chi_j=S_j^xS_{j+1}^y-S_j^yS_{j+1}^x$, and $\delta f_j=f_{j+1}-f_j$, the twist leaves the longitudinal exchange untouched and gives
\begin{equation}
 U_f^\dagger H U_f-H
 =J\sum_j[(\cos\delta f_j-1)B_j+\sin\delta f_j\,\chi_j].
 \label{eq:S-rotatedbond}
\end{equation}
For a translation-invariant target without spin current, $\langle\chi_j\rangle_0=0$, and with $C_\perp=\langle B_j\rangle_0$ this yields the exact relation
\begin{equation}
 \varepsilon_f(0)=JC_\perp\sum_j(\cos\delta f_j-1),
 \label{eq:S-injection}
\end{equation}
valid at any twist amplitude. Antiferromagnetic targets have $C_\perp<0$, so twisting neighboring spins away from their preferred relative angle costs energy, whereas a uniform rotation costs none.

For $f_m(j)=\alpha\sin(q_mj)$ with $q_m=2\pi m/L$ (excluding $q_m=0,\pi$, where the sine vanishes on all sites),
\begin{equation}
 \delta f_j=2\alpha\sin(q_m/2)\cos[q_m(j+1/2)],\qquad
 \sum_j(\delta f_j)^2=2\alpha^2L\sin^2(q_m/2).
 \label{eq:S-bondgradient}
\end{equation}
Expanding the cosine for small bond gradients, $|\alpha\sin(q_m/2)|\ll1$, gives
\begin{equation}
 \varepsilon_m(0)=-JC_\perp\alpha^2L\sin^2(q_m/2)
 +O\!\left(JL\alpha^4\sin^4(q_m/2)\right)
 \simeq-\frac{JC_\perp\pi^2\alpha^2m^2}{L},
 \label{eq:S-injection-smallq}
\end{equation}
where the last form uses $q_m\ll1$. Higher harmonics inject more energy because their phase gradients are steeper. Note that a small bond gradient does not require a weak overall displacement: at fixed $\alpha$ and $m$ the state remains finitely displaced as $L$ grows.

The twist also imposes an exact lattice momentum rule. Let $T$ translate by one site and $d=\gcd(m,L)$. Since $f_m(j+L/d)=f_m(j)$, $U_m$ commutes with $T^{L/d}$. If the target has momentum $k_0=2\pi\nu_0/L$, the twisted state therefore has weight only in momentum sectors $k_\nu=2\pi\nu/L$ with
\begin{equation}
 \ee^{\ii(k_\nu-k_0)L/d}=1,
 \qquad \nu-\nu_0=0\pmod d,
 \label{eq:S-subtranslation}
\end{equation}
at any amplitude and at all later times, since the evolution conserves momentum. At linear order the sine transfers momentum $\pm q_m$, and higher orders generate integer combinations of it. For $m=2$ on an even ring all odd relative momenta are excluded, while $m=1$ imposes no restriction; when $m$ and $L$ are coprime, the rule is empty. The lattice rule is weaker than the Gaussian grade selection of Eq.~\eqref{eq:S-gradeweights}: it removes entire momentum sectors but not every slow excitation within an allowed sector. Adding the $m=1$ component to the hot profile breaks its half-translation symmetry and reopens the odd sectors.

In the critical regime $-1<\Delta<1$ at zero magnetization, the Bethe ansatz gives~\cite{Haldane1980,LutherPeschel1975}
\begin{equation}
 \Delta=\cos\mu,\qquad
 K=\frac{\pi}{2(\pi-\mu)},\qquad
 v=\frac{J\pi\sin\mu}{2\mu},
 \label{eq:S-bethe}
\end{equation}
and Table~\ref{tab:S-criticalmethods} lists $K$ and $v/J$ for the anisotropies studied here. The Heisenberg point $\Delta=1$ is the limit $\mu\to0$, where a marginally irrelevant umklapp operator produces logarithmic corrections~\cite{Lukyanov1998}. At fixed $m$, $\omega_m\propto 1/L$ and $A_m=O(1)$, so the crossing times scale linearly with $L$. The limit $\Delta\to-1$ is singular in this description: the velocity vanishes and the dispersion becomes quadratic (Sec.~\ref{sec:fm}).

\begin{table}[tb]
 \centering
 \caption{\label{tab:S-criticalmethods}Luttinger parameter $K$ and spin velocity $v$ from Eq.~\eqref{eq:S-bethe} at the anisotropies studied (at $\Delta=1$, the limit $\mu\to0$), together with the parameters of the critical-chain MPS calculations.}
 \setlength{\tabcolsep}{9pt}
 \begin{tabular}{lcccccc}
 \toprule
 & \multicolumn{2}{c}{Bethe ansatz} & \multicolumn{4}{c}{MPS calculation}\\
 \cmidrule(lr){2-3}\cmidrule(lr){4-7}
 $\Delta$ & $K$ & $v/J$ & $L$ & $\delta\tau J$ & $\chi_{\max}$ & $\var(H)/J^2$\\
 \midrule
 $0$ & $1$ & $1$ & 40 & 0.02, 0.01 & 413 & $4\times10^{-8}$\\
 $1/2$ & $3/4$ & $3\sqrt3/4\approx1.30$ & 40 & 0.02, 0.01 & 455 & $6\times10^{-8}$\\
 $\cos(\pi/4)$ & $2/3$ & $\sqrt2\approx1.41$ & 40 & 0.02, 0.01 & 465 & $7\times10^{-8}$\\
 $1$ & $1/2$ & $\pi/2\approx1.57$ & 40 & 0.02, 0.01 & 472 & $9\times10^{-8}$\\
 \bottomrule
 \end{tabular}
\end{table}

The Gaussian initial energy $\omega_mA_m=\pi vK\alpha^2m^2/L$ has the same $m^2/L$ form as Eq.~\eqref{eq:S-injection-smallq}, but not necessarily the same coefficient. At $\Delta=0$ the two coincide, since $C_\perp=-1/\pi$ and $vK=J$: for free fermions, all particle--hole pairs created by the twist lie on the nearly linear branch close to the Fermi points. At $\Delta=1$, $C_\perp=\tfrac23(\tfrac14-\ln2)$ and the lattice coefficient exceeds the Gaussian one by about $18\%$. The exact injected energy is the first spectral moment $\sum_{\nu>0}e_\nu p_{\nu,m}$, and in the interacting chain a small population of higher-energy states contributes to it at the same order in $1/L$. These components decay on times short compared with $L/v$, so the crossings at $\tau\sim L/v$ are governed by the oscillator populations even though the initial energies differ. The relation~\eqref{eq:S-injection} also remains valid in the gapped regime, where no compact-boson description exists.

The MPS calculation are performed as follows. DMRG prepares the ground state of the periodic chain, and two-site TDVP evolves the two twisted copies independently under the same Hamiltonian, renormalizing the state after each step~\cite{White1992,HaegemanEtAl2016,FishmanEtAl2022}. The simulation parameters are listed in Table~\ref{tab:S-criticalmethods}. At $L=40$, both normalized differences cross zero once and remain negative up to the largest times computed, $\tau J/L=0.2$. As $\Delta$ increases from $0$ to $1$, the spin velocity grows and both crossings move to earlier times, while the shape of the curves changes little; the corresponding QMC trajectories at $L=80$ are shown in Fig.~\ref{fig:S-QMCbeta}. At $\Delta=0$, the chain maps to free fermions with antiperiodic boundary conditions in the half-filled sector of the even ring, and the MPS trajectories agree with the exact ones to within $10^{-6}$. Halving the time step changes the normalized differences and the crossing times by less than $10^{-6}$ in units of $L/J$. The crossing times also depend only weakly on how the sampled curves are interpolated near the zero, with changes below $4\times10^{-5}L/J$. All of these numerical uncertainties are more than an order of magnitude smaller than the finite-size deviations from the compact-boson prediction discussed in the Letter.

\subsection{Gapped N\'eel chains}
\label{sec:neel}
For a weak twist, the excited weights are given by the longitudinal dynamical structure factor $S^{zz}(q,\omega)=L^{-1}\sum_{\nu>0} |\langle\nu|S_q^z|\psi_0\rangle|^2\delta(\omega-e_\nu)$, where $S_q^z=\sum_j\ee^{-\ii qj}S_j^z$. Expanding $U_m=\ee^{\ii\alpha G_m}$ to second order and using the two Fourier components of the sine gives
\begin{equation}
 R_m(\tau)=\frac{\alpha^2L}{2}\int_0^\infty
 S^{zz}(q_m,\omega)\ee^{-2\omega\tau}\,d\omega
 +O(\alpha^4).
 \label{eq:S-response}
\end{equation}
In a gapped, magnetization-conserving phase the small-$q$ longitudinal weight scales as $q^2$. At fixed $m$ the initial infidelity is of order $\alpha^2/L$, much smaller than the $O(1)$ oscillator occupation of the critical chain. The relevant relaxation scale is the energy of the populated bulk excitations, not the smallest level splitting of the finite ring. This matters because of the two nearly degenerate N\'eel states. On the even rings used here, a reflection about the origin of the sine reverses $G_m$, and so does a global spin flip. Their product therefore commutes with $U_m$ while exchanging the two N\'eel configurations. The symmetric target lies in one eigenspace of this product, and the exponentially small doublet splitting is not populated by the twist.

Deep in the N\'eel phase, the transverse exchange $J/2$ acting on an antiparallel bond creates two domain walls at cost $J\Delta$. Second-order perturbation theory gives the energy $-J/(4\Delta)$ per bond and $C_\perp=-1/(2\Delta)+O(\Delta^{-3})$, so Eq.~\eqref{eq:S-injection-smallq} becomes
\begin{equation}
 \varepsilon_m(0)\simeq
 \frac{J\pi^2\alpha^2m^2}{2\Delta L},\qquad
 \overline e_m\equiv\frac{\varepsilon_m(0)}{I_m(0)}
 =\frac{\sum_{\nu>0}e_\nu p_{\nu,m}}
 {\sum_{\nu>0}p_{\nu,m}}.
 \label{eq:S-neelscale}
\end{equation}
The ratio $\overline e_m$ is the mean energy of the weight that the twist moves out of the target. By Eq.~\eqref{eq:S-cooling}, it fixes only the initial logarithmic decay rate of the infidelity, $-\partial_\tau\ln I_m|_{\tau=0}=2\overline e_m[1-I_m(0)]$. As the high-energy components decay, the relaxation becomes governed by the lowest populated excitations and their weights (Sec.~\ref{sec:spectral}), so a large $\overline e_m$ does not by itself imply fast relaxation at the times where a crossing can occur.

The $L=32$ data of Letter Fig.~3(a,b) show how the populations and the relaxation change with $\Delta$. Between $\Delta=1.2$ and $4$, the initial energy drops from $0.022J$ to $0.009J$ for the cold state and from $0.087J$ to $0.037J$ for the hot state, while the initial mean excitation energies rise from $0.38J$ to $3.4J$ and from $0.74J$ to $3.6J$, respectively. The subsequent relaxation is characterized by the effective decay rates $\Gamma_m(\tau)=-\tfrac12\partial_\tau\ln\varepsilon_m$, which approach the energy of the lowest populated excitation at late times. At $\Delta=1.2$ they settle by $\tau J\approx1$ at $\Gamma_c\approx0.34J$ and $\Gamma_h\approx0.66J$, so the hot state relaxes about 1.9 times faster than the cold one; at $\Delta=2$ the late-time ratio is about $1.5$. At $\Delta=4$ the effective rates of both states are still decreasing when the deviations become unresolvable, and over the whole resolvable window the hot state relaxes only 3--10\% faster than the cold one. As $\Delta$ increases, the two states thus relax at increasingly similar rates from decreasing initial deviations, which delays a possible crossing and reduces the signal. At $\Delta=4$ the hot state remains above the cold state in both energy and infidelity until the deviations of both states fall below the numerical precision of $10^{-8}$. Within this precision, $\Delta=4$ serves as a null case without reversal.

\subsection{The ferromagnetic endpoint and the dynamical exponent}
\label{sec:fm}
For even $L$, the staggered rotation $W=\prod_{j\,{\rm odd}}(2S_j^z)$ maps the $\Delta=-1$ XXZ chain onto the ferromagnet $H_F=-J\sum_j\bm S_j\cdot\bm S_{j+1}$ and commutes with $U_f$~\cite{AlbaHaqueLauchli2012}. In the sector with $n$ down spins the ground state is the Dicke state
\begin{equation}
 |D_n\rangle=\binom Ln^{-1/2}
 \sum_{j_1<\cdots<j_n}|j_1,\ldots,j_n\rangle
 \propto(S^-_{\rm tot})^n|F\rangle,
 \qquad S^z_{\rm tot}=L/2-n,
 \label{eq:S-Dicke}
\end{equation}
where $|F\rangle$ is fully polarized up. The ground multiplet has total spin $L/2$, and $|D_n\rangle$ is its unique member at fixed $n$~\cite{Dicke1954}; the target in the original frame is $W^\dagger|D_n\rangle$. In the fully polarized sectors $n=0,L$ the twist produces only a global phase. The elementary spin wave of the ferromagnet has energy
\begin{equation}
 e_m=J[1-\cos(2\pi m/L)]\simeq\frac{2\pi^2Jm^2}{L^2}.
 \label{eq:S-magnon}
\end{equation}

Permutation symmetry gives $\var(S_j^z)=n(L-n)/L^2$ and the off-site covariance $-n(L-n)/[L^2(L-1)]$. Together with $\sum_j\sin(q_mj)=0$ and $\sum_j\sin^2(q_mj)=L/2$, this yields
\begin{equation}
 \var_{D_n}(G_m)=\frac{n(L-n)}{2(L-1)},\qquad
 w_m=\alpha^2\frac{n(L-n)}{2(L-1)}+O(\alpha^4),
 \qquad \varepsilon_m(0)=e_mw_m+O(\alpha^4).
 \label{eq:S-Dickevar}
\end{equation}
The first-order state $G_m|D_n\rangle$ is a descendant of an exact one-spin-wave state, because $[H_F,S^-_{\rm tot}]=0$, and therefore has energy $e_m$ at any filling. In contrast to the compact boson, the leading excited weight $w_m$ does not depend on the harmonic index. Using the transverse bond expectation value $n(L-n)/[L(L-1)]$ in the ferromagnetic frame, Eq.~\eqref{eq:S-rotatedbond} gives the exact injected energy
\begin{equation}
 \varepsilon_m(0)=
 \frac{Jn(L-n)}{L(L-1)}\sum_j[1-\cos\delta f_j].
 \label{eq:S-FMinjection}
\end{equation}

When the weight created by a weak twist is concentrated at a single energy $e_m$, $\varepsilon_m\simeq e_mw_m\ee^{-2e_m\tau}$ and $I_m\simeq w_m\ee^{-2e_m\tau}$, and the crossing times are
\begin{equation}
 \tau_E^{\rm weak}=\frac{\ln(e_hw_h/e_cw_c)}{2(e_h-e_c)},
 \qquad \tau_I^{\rm weak}=
 \frac{\ln(w_h/w_c)}{2(e_h-e_c)}.
 \label{eq:S-weakroots}
\end{equation}
For $e_h>e_c$, an energy crossing requires $e_hw_h>e_cw_c$ and an infidelity crossing requires $w_h>w_c$. At equal amplitudes $w_h=w_c$ to leading order, so the ferromagnet shows an energy reversal without an infidelity reversal, as discussed in the Letter. With $e_m\propto m^2/L^2$, $\tau_E^{\rm weak}=L^2\ln4/(12\pi^2J)\approx0.0117L^2/J$ for $(m_h,m_c)=(2,1)$.

The one-magnon sector $n=1$ can be solved exactly at any twist amplitude. Since $\sum_j\sin(2\pi mj/L)=0$, the twist multiplies the amplitude of the flipped spin at site $j$ by $\ee^{-\ii\alpha\sin(2\pi mj/L)}$, so the twisted state is a superposition of one-magnon momentum states with weights and energies
\begin{equation}
 p_{m,\ell}
 =
 \left|
 \frac{1}{L}\sum_{j=0}^{L-1}
 \ee^{-\ii\alpha\sin(2\pi mj/L)}
 \ee^{-2\pi\ii\ell j/L}
 \right|^2,
 \qquad
 e_\ell=J[1-\cos(2\pi\ell/L)].
 \label{eq:S-oneweights}
\end{equation}
Inserted into Eqs.~\eqref{eq:S-Z}--\eqref{eq:S-ratios}, these give the exact finite-amplitude trajectories without any weak-twist expansion; the one-magnon crossing times in the inset of Letter Fig.~3(c) are obtained in this way. Both harmonics have the same target weight, $p_{m,0}\simeq\mathcal J_0(\alpha)^2$ with $\mathcal J_0$ the zeroth-order Bessel function, so the initial infidelities coincide and only the energy ordering can reverse. At $\alpha=1/2$ the energy crossing approaches $\tau_EJ/L^2\approx0.0121$ for large $L$, slightly above the weak-twist value.

In the half-filled sector the energy crossings occur at $\tau_EJ/L^2=0.0140$, $0.0128$, $0.0124$, and $0.0122$ for $L=8$, $12$, $16$, and $20$, with ED and MPS agreeing at $L=16$, as shown in the Letter Fig~3(c) insert. These values lie 2--4\% above Eq.~\eqref{eq:S-weakroots} evaluated with the exact magnon energies, and the difference decreases with $L$. Because $w_m\simeq\alpha^2L/8$ at half filling, a fixed amplitude $\alpha=1/2$ does not remain weak as $L$ grows, and a controlled thermodynamic limit requires $\alpha^2L\ll1$. More generally, a dispersion $e_m\propto m^zL^{-z}$ gives $\tau_\times\propto L^z$ whenever the weight ratio $w_h/w_c$ approaches a size-independent limit and the corresponding crossing exists. The dynamical exponent thus sets the time scale of the reversal, while its existence is decided by the spectral weights~\cite{ChangEtAl2026ITME}.

\subsection{Finite-width ladders}
\label{sec:ladders}
On an $L_x\times L_y$ square ladder we consider
\begin{equation}
 H=J\sum_{\langle ij\rangle}
 (S_i^xS_j^x+S_i^yS_j^y+\Delta S_i^zS_j^z),\qquad
 f_m(x,y)=\alpha\sin(2\pi mx/L_x),
 \label{eq:S-ladder}
\end{equation}
with periodic boundary conditions along the legs and open boundaries across them. The twist is uniform on each rung. All invariances of Sec.~\ref{sec:twist} carry over, but the relaxation spectrum is no longer that of a single oscillator.

For the $16\times2$ ladder of Letter Fig.~3(d), with $\Delta=\alpha=1/2$ in the $S^z_{\rm tot}=0$ sector, the target variance is $1.5\times10^{-9}J^2$. Two-site TDVP with truncation threshold $10^{-11}$ gives the infidelity crossing at $\tau_IJ\approx1.16$ and the energy crossing at $\tau_EJ\approx1.71$, with no recrossing up to $\tau J=4$. Halving the time step from $0.05$ to $0.025$ changes the curves by less than $4\times10^{-8}$ in both energy (units of $J$) and infidelity, and the bond dimension stays below its cap.

Smaller ladders allow a direct comparison with ED and a scan of the anisotropy. At $\Delta=1/2$ both observables reverse on the $8\times2$ and $6\times3$ ladders, with $(\tau_EJ,\tau_IJ)\approx(1.02,0.68)$ and $(1.06,0.74)$, respectively. At $\Delta=1$ the two-leg ladder still reverses in both observables, whereas the three-leg ladder shows only a transient energy reversal for $2.45\lesssim\tau J\lesssim3.0$. At $\Delta=2$ neither geometry reverses up to $\tau J=4$. ED and MPS agree closely for these sizes, and we count a sign change only when the difference exceeds ten times their discrepancy. These results show that the twist preparation carries over to ladders, but that a reversal depends on the spectrum of the populated modes and is not guaranteed. Since both the length and the width change between geometries, they do not constitute a width extrapolation or a two-dimensional limit.

\section{Directed-loop quantum Monte Carlo}
\label{sec:qmc}
\subsection{Stochastic series expansion and directed loops}
We simulate even periodic rings at zero field for $\Delta=0,1/2,\cos(\pi/4),1$ using the standard stochastic series expansion (SSE) with diagonal and directed-loop updates~\cite{Sandvik1999,SyljuasenSandvik2002,Sandvik2010}, and give here only the model-specific ingredients. A Marshall rotation on one sublattice changes the sign of the transverse exchange and commutes with the diagonal twist. The rotated Hamiltonian reads
\begin{equation}
 \widetilde H=\frac{JL}{4}-\sum_b(D_b+O_b),\qquad
 D_b=J(1/4-\Delta S_i^zS_j^z),\qquad
 O_b=\frac J2(S_i^+S_j^-+S_i^-S_j^+),
 \label{eq:S-SSEdecomp}
\end{equation}
with nonzero bond matrix elements
\begin{equation}
 w_p=\frac J4(1-\Delta),\qquad
 w_a=\frac J4(1+\Delta),\qquad w_o=\frac J2
 \label{eq:S-vertices}
\end{equation}
for parallel diagonal, antiparallel diagonal, and exchange vertices. All are nonnegative for $0\leq\Delta\leq1$, and the diagonal updates use these anisotropic weights. For the three vertices connected by a loop move, ordered as $(p,a,o)$, the symmetric solution of the directed-loop equations~\cite{SyljuasenSandvik2002} is
\begin{equation}
 \mathsf a=
 \begin{pmatrix}0&0&w_p\\0&0&w_a\\w_p&w_a&0\end{pmatrix},
 \qquad P_{ij}=a_{ij}/w_i.
 \label{eq:S-flows}
\end{equation}
Because $w_p+w_a=w_o$, every row sums to the corresponding vertex weight, and $w_iP_{ij}=w_jP_{ji}$ ensures detailed balance. A diagonal vertex always turns into an exchange vertex, and an exchange vertex turns into the parallel or antiparallel diagonal vertex with probabilities $(1-\Delta)/2$ and $(1+\Delta)/2$. Bounces are absent throughout $0\leq\Delta\leq1$; at $\Delta=1$ the parallel vertex has zero weight and drops out.

Loops that wind around the imaginary-time direction change the total magnetization, so the simulation samples the full zero-field thermal ensemble rather than a fixed-magnetization sector. All weights are positive and there is no sign problem; the twist enters only as a complex phase in the measured observable.

\subsection{Thermal twist correlator and projection}
With $Z_\beta=\Tr\ee^{-\beta H}$, we measure the thermal twist correlator and the twist expectation value
\begin{equation}
 C_f(s,\beta)=\frac{\Tr[\ee^{-(\beta-s)H}U_f^\dagger\ee^{-sH}U_f]}{Z_\beta},
 \qquad z_f(\beta)=\frac{\Tr[\ee^{-\beta H}U_f]}{Z_\beta}.
 \label{eq:S-thermal}
\end{equation}
At $s=2\tau$ these define the finite-temperature estimators
\begin{equation}
 \varepsilon_f^{(\beta)}=-\frac{C_f'}{C_f},\qquad
 [\sigma_f^{(\beta)}]^2=\frac{C_f''}{C_f}-\left(\frac{C_f'}{C_f}\right)^2,
 \qquad I_f^{(\beta)}=1-\frac{|z_f|^2}{C_f},
 \label{eq:S-qmcproxies}
\end{equation}
where primes denote derivatives with respect to $s$ at fixed $\beta$. When both $\beta-2\tau$ and $\beta$ are large compared with the inverse finite-size gap, $C_f(2\tau,\beta)\to Z_f(2\tau)$ and $z_f\to\langle\psi_0|U_f|\psi_0\rangle$, and Eq.~\eqref{eq:S-qmcproxies} reduces to the pure-state observables of Eq.~\eqref{eq:S-observables}. This is the same projection principle as in projector QMC, here applied to both ends of the imaginary-time interval. The slowest thermal contamination comes from the lowest excited states, which for $0\leq\Delta<1$ carry $S^z_{\rm tot}=\pm1$ and lie $\pi v/(2KL)$ above the ground state. At $\beta J=4L$ their Boltzmann factor is about $2\times10^{-3}$ at $\Delta=0$ and drops below $10^{-8}$ at $\Delta=1$. We therefore check projection explicitly by comparing three temperatures (Sec.~\ref{sec:qmc-production}).

\subsection{Measuring the twist correlator in continuous imaginary time}
At slot $p$ of the operator string, whose length $M$ includes identity operators, and with $\sigma_j(p)=\pm1$, we record the twist phase and its circular autocorrelation,
\begin{equation}
 u_f(p)=\exp\!\left[\frac{\ii}{2}\sum_jf_j\sigma_j(p)\right],
 \qquad
 A_k[f]=\frac1M\sum_{p=0}^{M-1}u_f^*(p+k)u_f(p),
 \label{eq:S-phase}
\end{equation}
with indices taken modulo $M$. Both harmonics are measured on the same configurations, and the full array $A_k$ is computed with a fast Fourier transform. Distributing the $M$ slots over $[0,\beta)$ gives the binomial kernel
\begin{equation}
 C_f(s,\beta)=\left\langle\sum_{k=0}^M
 B_{M,k}(s/\beta)A_k[f]\right\rangle,\qquad
 B_{n,k}(x)=\binom nkx^k(1-x)^{n-k},
 \label{eq:S-bernstein}
\end{equation}
which yields correlations at continuous imaginary time without a Trotter error~\cite{DorneichTroyer2001}. The derivatives follow from differentiating the kernel rather than from finite differences of noisy data:
\begin{align}
 C_f'(s)&=\left\langle\frac M\beta\sum_{k=0}^{M-1}
 B_{M-1,k}(s/\beta)(A_{k+1}-A_k)\right\rangle,\\
 C_f''(s)&=\left\langle\frac{M(M-1)}{\beta^2}
 \sum_{k=0}^{M-2}B_{M-2,k}(s/\beta)
 (A_{k+2}-2A_{k+1}+A_k)\right\rangle.
 \label{eq:S-bernsteinderivatives}
\end{align}
We average $C,C',C'',z$ first and form the ratios of Eq.~\eqref{eq:S-qmcproxies} afterwards. The imaginary parts of these averages vanish by symmetry and serve as a check of the sampling.

At $s=0$ only the first difference $A_1-A_0$ contributes to $C_f'$. Because the twist phase changes only where an exchange operator acts, the real part of this term reduces, configuration by configuration, to the number $n_{o,b}$ of exchange operators on bond $b$,
\begin{equation}
 \varepsilon_f^{(\beta)}(0)=
 \sum_b[1-\cos(f_{b+1}-f_b)]\frac{\langle n_{o,b}\rangle}{\beta},
 \label{eq:S-QMCinjection}
\end{equation}
which is the thermal counterpart of Eq.~\eqref{eq:S-injection}.

\subsection{Benchmarks against MPS}
\label{sec:qmc-benchmarks}

\begin{figure}[tb]
 \centering
 \includegraphics[width=\textwidth]{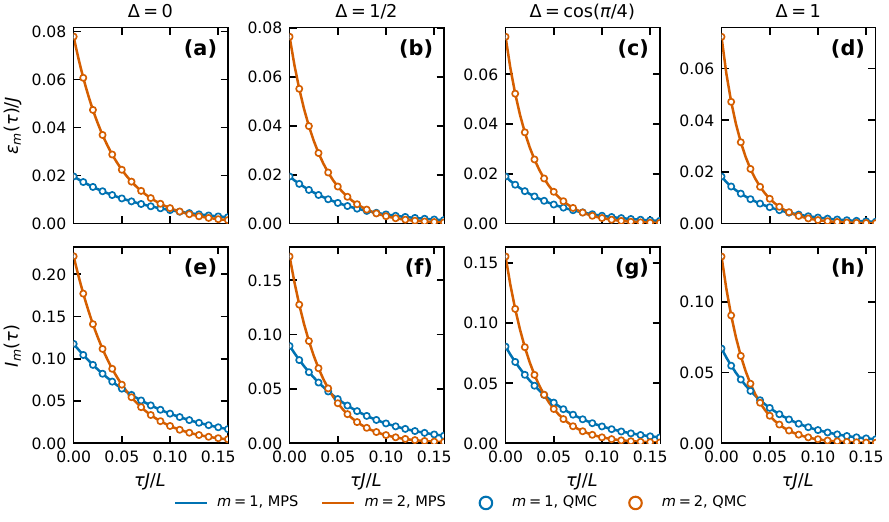}
 \caption{\label{fig:S-QMCvalidation}QMC at $\beta J=4L$ (open circles) compared with zero-temperature MPS (lines) for the same chain, $L=40$ and $\alpha=1/2$. Upper panels: excess energy; lower panels: target infidelity, for the cold $m=1$ (blue) and hot $m=2$ (orange) states. Every fourth QMC time point is shown; statistical errors are smaller than the symbols.}
\end{figure}

To test projection and the large-$L$ dynamics, we compare QMC at $\beta J=4L$ with the MPS ground-state trajectories at the same size, $L=40$. Figure~\ref{fig:S-QMCvalidation} shows the individual trajectories: the hot state starts with about four times the energy and twice the infidelity of the cold state, and overtakes it within the time window. The QMC data reproduce the MPS curves to within about $10^{-3}$ of their initial values at all times. The deviations are consistent with the statistical errors, which are strongly correlated between neighboring times, and the eight crossing times agree with MPS within $2.5$ standard deviations.

\subsection{Large-scale results at \texorpdfstring{$L=80$}{L=80}}
\label{sec:qmc-production}
The production runs cover $L=80$ and $\beta J=4L$. Observables are evaluated at $0\leq\tau J/L\leq0.16$ in steps of $0.0025$. Statistical errors are obtained by a joint bootstrap over the independent chains. Each resample selects the same chains for both profiles, all times, and all of $C,C',C'',z$, which keeps their correlations in the nonlinear ratios and in the hot-minus-cold differences.

\begin{table}[tb]
 \centering
\caption{\label{tab:S-QMCroots}QMC crossing times at $L=80$, $\beta J=4L$, and $\alpha=1/2$, used in Letter Fig.~2. The numbers in parentheses are statistical errors in the last digits. The last two columns give the compact-boson values of Eq.~\eqref{eq:S-equalroots} with the Bethe-ansatz velocity.}
 \begin{tabular}{lcccc}
 \toprule
 $\Delta$&$\tau_EJ/L$&$\tau_IJ/L$&$\tau_E^{\rm CB}J/L$&$\tau_I^{\rm CB}J/L$\\
 \midrule
 $0$&0.1108(7)&0.0552(3)&0.1103&0.0552\\
 $1/2$&0.0857(6)&0.0426(2)&0.0849&0.0425\\
 $\cos(\pi/4)$&0.0779(4)&0.0391(2)&0.0780&0.0390\\
 $1$&0.0707(4)&0.0359(2)&0.0702&0.0351\\
 \bottomrule
 \end{tabular}
\end{table}

\begin{figure}[tb]
 \centering
 \includegraphics[width=\textwidth]{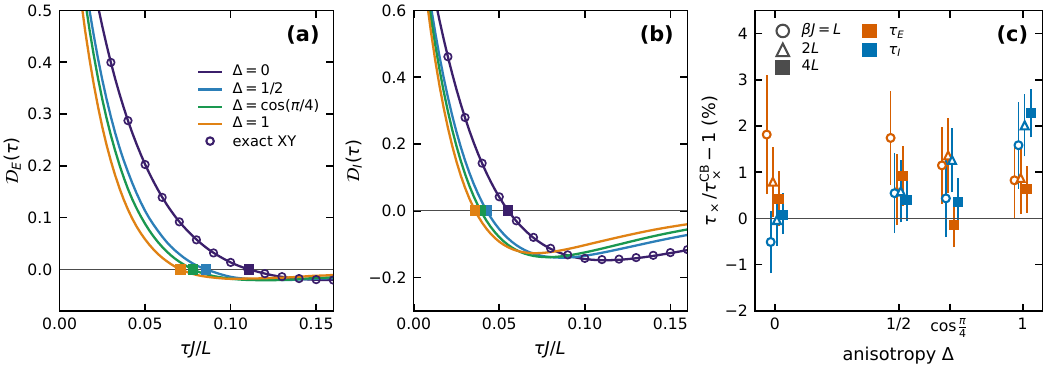}
\caption{\label{fig:S-QMCbeta}QMC at $L=80$. (a) Normalized energy difference and (b) normalized infidelity difference at $\beta J=4L$; the statistical errors (shaded) are narrower than the lines. Squares mark the crossing times, and open circles are the exact free-fermion results for the XY chain ($\Delta=0$) at the same size. (c) Relative deviation of the crossing times from the compact-boson values, for $\tau_E$ (orange) and $\tau_I$ (blue) at $\beta J=L$ (circles), $2L$ (triangles), and $4L$ (filled squares); symbols are displaced horizontally for clarity.}
\end{figure}

Both reversals are present in every bootstrap resample, and the positive initial and negative late differences are resolved by more than two standard errors. Table~\ref{tab:S-QMCroots} and Fig.~\ref{fig:S-QMCbeta} summarize the results. At $\Delta=0$ the QMC trajectories coincide with the exact free-fermion trajectories of the same $L=80$ chain [Fig.~\ref{fig:S-QMCbeta}(a,b)]. This comparison tests the full chain of QMC approximations, including thermal projection, at the size used in the Letter. For all anisotropies, the trajectories closely resemble the $L=40$ MPS results when time is measured in units of $L$, in line with the linear scaling of the crossing times. The crossing times agree with the compact-boson values to within about $1\%$ [Fig.~\ref{fig:S-QMCbeta}(c)]. Results at $\beta J=2L$ and $4L$ agree within 1.6 standard errors, and no systematic temperature dependence is resolved. The $\beta J=L$ data have larger errors but are compatible as well. The comparison with ED at $L=16$ and MPS at $L=40$ in Letter Fig.~2(b) reflects finite-size effects instead of a benchmark comparing between the methods.

\section{Variational quantum imaginary-time evolution}
\label{sec:vqite}
\subsection{Circuit and measurements}
A state-preparation circuit $W_0$ maps the initialized register to the common target, $W_0|0\cdots0\rangle=|\psi_0\rangle$. Since the target has $S^z_{\rm tot}=0$, this step cannot be performed with magnetization-conserving gates alone, and in our simulations the target is supplied directly; we do not address an efficient hardware construction of $W_0$. With $S_j^z=Z_j/2$ and $R_z(\vartheta)=\exp(-\ii\vartheta Z/2)$,
\begin{equation}
 U_m=\prod_jR_{z,j}[-\alpha\sin(2\pi mj/L)],\qquad
 |\psi(\bm\theta)\rangle=V_D(\bm\theta)U_m|\psi_0\rangle.
 \label{eq:S-circuit}
\end{equation}
The twist is a single layer of single-qubit rotations. Each of the $D$ brick-wall layers of the variational circuit $V_D$ combines $R_{XY}(\vartheta)=\exp[-\ii\vartheta(XX+YY)/4]$, $R_{ZZ}(\vartheta)=\exp(-\ii\vartheta ZZ/4)$, and local $R_z$ gates, all of which conserve the magnetization. The angles of gate family $g$ in layer $d$ vary smoothly in space,
\begin{equation}
 \theta_{g,j,d}=a_{g,0,d}+
 \sum_{r=1}^{M_a}
 [a_{g,r,d}\sin(2\pi rj/L)+b_{g,r,d}\cos(2\pi rj/L)],
 \label{eq:S-fourierangles}
\end{equation}
with a spatial bandwidth $M_a$. For the hot $m=2$ profile we restrict the angles to the half-translation-symmetric sector of the state, which reduces the number of parameters at $D=6$, $M_a=2$ from 90 to 54; the profiles with slow-mode admixture use the full parametrization.

The energy follows from $XX$, $YY$, and $ZZ$ bond correlators measured in three bases. The metric and gradient can be obtained from ancilla-assisted or parameter-shift measurements~\cite{McArdleEtAl2019,YuanEtAl2019}. If the inverse preparation $W_0^\dagger$ is available, the target fidelity is the probability of the all-zero outcome after applying it. In an experiment, the local energy is thus considerably easier to access than the target fidelity. Here VQITE updates a fixed parametrized circuit through the McLachlan equations, whereas local QITE constructs a unitary approximation to each imaginary-time step on a spatial domain~\cite{MottaEtAl2020}.

\subsection{McLachlan projection and tangent residual}
After removing the global phase direction, $|D_a\psi\rangle=(1-|\psi\rangle\langle\psi|) |\partial_a\psi\rangle$, McLachlan's principle chooses the parameter velocity that best approximates the exact imaginary-time step,
\begin{equation}
 \min_{\dot{\bm\theta}}\left\|\sum_a\dot\theta_a|D_a\psi\rangle
 +(H-E)|\psi\rangle\right\|^2,
 \qquad E=\langle\psi|H|\psi\rangle,
 \label{eq:S-McLfunctional}
\end{equation}
which leads to
\begin{equation}
 G\dot{\bm\theta}=-\bm C,\qquad
 G_{ab}=\Re\langle D_a\psi|D_b\psi\rangle,\qquad
 C_a=\Re\langle D_a\psi|(H-E)|\psi\rangle
 =\tfrac12\partial_aE.
 \label{eq:S-McL}
\end{equation}
This projects the exact flow onto the real tangent space of the ansatz~\cite{McArdleEtAl2019,YuanEtAl2019}. For a given parameter velocity, the deviation of the update from the exact imaginary-time velocity $-(H-E)|\psi\rangle$, whose norm is $\sigma_H$, is measured by the physical residual
\begin{equation}
 \mathcal R_\perp^2=\sigma_H^2+
 2\dot{\bm\theta}^{T}\bm C+
 \dot{\bm\theta}^{T}G\dot{\bm\theta}.
 \label{eq:S-residual}
\end{equation}
For the unregularized pseudoinverse solution it equals $\sigma_H^2-\bm C^TG^+\bm C$, the squared norm of the component of the exact velocity outside the tangent space. We solve Eq.~\eqref{eq:S-McL} with a truncated pseudoinverse (relative singular-value threshold $10^{-6}$) and fourth-order Runge--Kutta integration, and evaluate Eq.~\eqref{eq:S-residual} with the velocity actually applied. The resulting ratio $\mathcal R_\perp/\sigma_H$ is the normalized physical residual of the regularized update. Besides the component outside the tangent space, it contains the tangent components discarded by the regularization, and it is therefore never smaller than the unregularized projection residual. It is an instantaneous measure of the update error, not a bound on the accumulated state error. A small residual of the linear equation $G\dot{\bm\theta}+\bm C$ does not imply a small $\mathcal R_\perp$, because the former cannot detect directions perpendicular to the tangent space.

\subsection{Accuracy of the twelve-qubit simulations}
The circuits are simulated with TensorCircuit and the JAX backend in double precision~\cite{ZhangEtAlTensorCircuit2023}. The benchmarks use $L=12$, $\Delta=\alpha=1/2$, $M_a=2$, and $\delta\tau J=0.01$; the exact reference is computed for the same Hamiltonian, sector, and twist. Crossing-time errors are quoted as $\eta_\times=100(\tau_{\times,\rm VQITE}-\tau_{\times,\rm exact})/\tau_{\times,\rm exact}$. As reported in the Letter, increasing the depth from $D=2$ to $6$ reduces $\eta_E$ from $0.6\%$ to $0.3\%$ and $\eta_I$ from $1.1\%$ to $0.6\%$, and VQITE reproduces the delay caused by a slow-mode admixture $\gamma=0.5$.

\begin{figure}[tb]
 \centering
 \includegraphics[width=.78\textwidth]{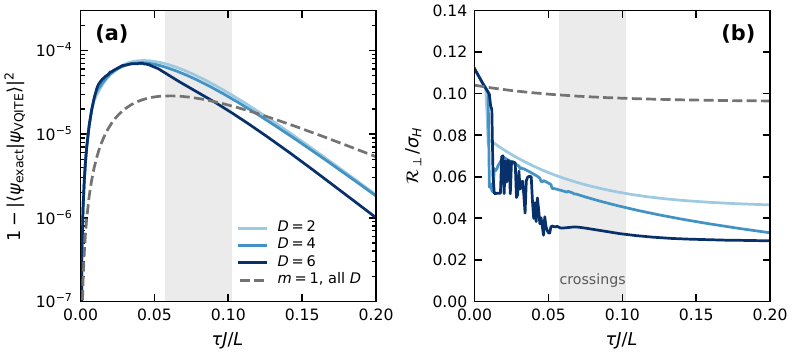}
 \caption{\label{fig:S-vqite}Accuracy of VQITE at $L=12$, $\Delta=\alpha=1/2$, $M_a=2$, for the hot $m=2$ profile at depths $D=2,4,6$ and for the cold $m=1$ profile (identical for all $D$). (a) Infidelity between the VQITE state and the exact imaginary-time state. (b) Normalized physical residual of the regularized update, $\mathcal R_\perp/\sigma_H$ from Eq.~\eqref{eq:S-residual}; the step-like features at early times occur when singular values of the metric cross the regularization threshold. The shaded interval lies between the exact infidelity and energy crossing times.}
\end{figure}

The agreement of the crossing times reflects an accurate state trajectory [Fig.~\ref{fig:S-vqite}(a)]. The infidelity to the exact imaginary-time state stays below $10^{-4}$ throughout the evolution. It grows during the initial fast relaxation, when the state changes most rapidly, and decreases again as both states approach the ground state. The normalized residual of the regularized update [Fig.~\ref{fig:S-vqite}(b)] starts at about $0.1$ for both profiles, i.e., the applied update initially deviates from the exact imaginary-time velocity by about $10\%$ of its norm. Because this residual bounds the unregularized projection residual from above, at least $98\%$ of the squared norm of the exact velocity lies within the tangent space throughout the evolution. For the hot state the residual decreases with time and with depth, to $0.03$--$0.05$ at late times, consistent with the smaller crossing-time errors at larger $D$. For the cold state the residual stays near $0.1$, and both the residual and the trajectory infidelity are identical for $D=2,4,6$; along this trajectory the additional layers do not enlarge the set of update directions retained by the regularization. Halving the time step changes the crossing times by less than $10^{-4}L/J$.

\section{Relation to monitored dynamics}
\label{sec:nonhermitian}
\subsection{Conditional no-jump evolution}
\label{sec:nojump}
In a monitored system described by a Lindblad equation, a trajectory without quantum jumps evolves under the effective Hamiltonian $H_{\rm eff}=H_c-\ii\sum_\mu J_\mu^\dagger J_\mu/2$~\cite{DalibardCastinMolmer1992}. Within the chosen sector, $H_0=H-E_0\geq0$, and the choice $H_c=0$ with a single jump operator $J_{\rm glob}=\sqrt{2H_0}$ gives
\begin{equation}
 H_{\rm eff}=-\ii H_0,\qquad
 |\widetilde\psi_f(t)\rangle=\ee^{-tH_0}|\phi_f\rangle,
 \qquad P_{{\rm nj},f}(t)=Z_f(2t).
 \label{eq:S-nj}
\end{equation}
After normalization this is exactly the imaginary-time state at $\tau=t$. The probability $P_{{\rm nj},f}$ of no jump up to time $t$ is not the target fidelity. Its hazard, the jump rate conditioned on no earlier jump, is
\begin{equation}
 \lambda_f=-\partial_t\ln P_{{\rm nj},f}=2\varepsilon_f.
 \label{eq:S-hazard}
\end{equation}
An energy reversal is therefore a crossing of conditional jump rates; the survival probabilities themselves need not cross.

Local jump operators give the same normalized trajectory but a different survival probability. For $-1\leq\Delta\leq1$ let $k_j=h_j+c_\Delta\mathbb1$ with $c_\Delta=J(2+\Delta)/4$, where $h_j$ is the XXZ bond Hamiltonian. The eigenvalues of $k_j$ are $0$, $J$, and $J(1+\Delta)/2$ (twice), so $J_j=\sqrt{2\kappa k_j}$ is a well-defined two-site jump operator. The resulting $H_{\rm eff}^{\rm loc}=-\ii\kappa(H+c_\Delta L)$ differs from $-\ii\kappa H_0$ by a constant, which drops out of the normalized state, so that $\tau=\kappa t$, but enters the survival probability:
\begin{equation}
 P_{{\rm nj},f}^{\rm loc}(t)=
 \ee^{-2\kappa t(E_0+c_\Delta L)}Z_f(2\kappa t),\qquad
 \lambda_h^{\rm loc}-\lambda_c^{\rm loc}
 =2\kappa(\varepsilon_h-\varepsilon_c).
 \label{eq:S-localnj}
\end{equation}
At $\Delta=1$, $k_j/J$ is the projector onto the bond triplet. For the critical chain the offset $E_0+c_\Delta L$ is extensive, so at the crossing time $\tau\propto L$ one has $-\ln P_{\rm nj}^{\rm loc}=O(L^2)$: local monitoring reproduces the dynamics, but only with exponentially rare postselection. At the frustration-free point $\Delta=-1$ the offset vanishes in the fixed-sector target. Once the jump terms are retained, the dynamics changes qualitatively. The Hermitian global jump dephases the state without cooling it, and the Hermitian bond jumps generate a unital channel~\cite{MingantiEtAl2019}. A dissipative analogue of the present construction therefore requires a separate analysis of the stationary state and the slow Liouvillian modes, including their symmetry structure~\cite{WeiEtAl2026SymmetryMpemba}.

\subsection{Coherent states under linear loss}
\label{sec:S-coherentbridge}
The Gaussian limit offers an exception in which jumps do not spoil the correspondence. Consider positive-frequency standing-wave oscillators, with charge and winding zero modes fixed, subject to independent zero-temperature loss:
\begin{equation}
 H_{\rm osc}=\sum_m\omega_m b_m^\dagger b_m,\qquad
 \dot\rho=-\ii[H_{\rm osc},\rho]+\sum_m2\gamma_m\mathcal D[b_m]\rho,
 \label{eq:S-bosonloss}
\end{equation}
where $\mathcal D[b]\rho=b\rho b^\dagger-\{b^\dagger b,\rho\}/2$ and $\gamma_m>0$ is the damping rate. Such linear loss maps coherent states onto coherent states under the full, unconditional evolution~\cite{Longhi2024BosonicMpemba}. For a coherent input $|\boldsymbol\zeta(0)\rangle$,
\begin{equation}
 \rho(t)=|\boldsymbol\zeta(t)\rangle\langle\boldsymbol\zeta(t)|,
 \qquad \zeta_m(t)=\zeta_m(0)\ee^{-(\gamma_m+\ii\omega_m)t}.
 \label{eq:S-coherentloss}
\end{equation}
Each jump acts on the instantaneous coherent state only by a scalar factor, so all measurement records lead to the same normalized state. If the ratio $\gamma_m/\omega_m=\kappa$ is the same for all occupied modes, this state coincides with the normalized $\ee^{-(\kappa+\ii)tH_{\rm osc}}|\boldsymbol\zeta(0)\rangle$, i.e., with imaginary-time evolution combined with free rotation. For mode-dependent damping the slow-mode filtering, and hence the possibility of a reversal, persists, but the exact correspondence is lost. This is a statement about effective oscillators: a microscopic reservoir for the XXZ chain must supply the noise, conserved sectors, and stationary state independently, and thermal noise or interactions generally destroy the pure coherent-state manifold.

\end{document}